\documentclass[aps,superscriptaddress,amsmath,twocolumn,amssymb,showpacs]{revtex4}
\usepackage{epsfig}

\begin{document}

\title{Numerical study of a fragile three-dimensional kinetically
constrained model}

\author{Ludovic Berthier}

\affiliation{Laboratoire des Verres UMR 5587, Universit\'e Montpellier
II and CNRS, 34095 Montpellier, France}

\author{Juan P. Garrahan}

\affiliation{School of Physics and Astronomy, University of
Nottingham, Nottingham, NG7 2RD, UK}

\date{\today}

\begin{abstract}
We numerically study the three-dimensional generalization of the
kinetically constrained East model, the North-or-East-or-Front (NEF)
model.  We characterize the equilibrium behaviour of the NEF model in
detail, measuring the temperature dependence of several quantities:
$\alpha$-relaxation time, distributions of relaxation times, dynamic
susceptibility, dynamic correlation length, and four-point
susceptibility.  We show that the NEF model describes quantitatively
experimental observations over an exceptionally wide range of
timescales.  We illustrate this by fitting experimental data obtained
both in the mildly supercooled regime by optical Kerr effect, and
close to the glass transition by dielectric spectroscopy.
\end{abstract}


\maketitle

\section{Introduction}

Supercooled liquids can be studied using theoretical models where the
density field dynamics is replaced by a coarse-grained mobility field
evolving with simple empirical rules based on the idea of dynamic
facilitation~\cite{reviewkcm}.  Kinetically constrained models (KCM),
such as the Fredrickson-Andersen (FA) model \cite{fa} and the East
model \cite{east,Sollich-Evans}, have been shown to reproduce well the
phenomenology of supercooled liquids.  An important feature of these
models is that their dynamics becomes spatially
correlated~\cite{harrowell,garrahan-chandler}, i.e., growing
timescales are accompanied by growing dynamical lengthscales, giving
rise to dynamic heterogeneity as observed in experiments and
simulations~\cite{DHreviews1,DHreviews2,DHreviews3,DHreviews4}.

In this paper, we study by means of extensive numerical simulations
the three-dimensional generalization of the East model~\cite{east},
the North-or-East-or-Front (NEF) model \cite{NEother}.  This model is
a realization of the `arrow' model of Ref.\ \cite{pnas}, but with an
externally imposed directional preference while in the arrow model
excitations carry an orientation which locally determines the
directionality of the dynamics, making the system isotropic~\cite{pnas}.  
These models were shown to reproduce the
dynamic behaviour of a wide range of fragile materials~\cite{pnas}.
The present study is complementary to that of
Refs.~\cite{steve1,steve2} that considered systems with fully
isotropic dynamic facilitation, such as the three-dimensional FA
model, which describes strong materials.

The principal aims of this work are the following:

{\em (i) Quantitative characterization of the fragile limit}.  We
report numerical results for a wide range of dynamic observables in
the NEF model, describing in detail distributions of the relevant
timescales and lengthscales.  We also confirm some of the predictions
for the arrow model of Ref.~\cite{pnas}.

{\em (ii) Comparison to the strong limit}.  We contrast our results
for the NEF model to those obtained in Refs.~\cite{steve1,steve2}
which dealt with the strong limit of isotropic facilitation, such as
the three-dimensional FA model.

{\em (iii) Comparison to experimental data}.  We use the NEF model
results to fit experimental data over a wide range of relaxation
times, covering both the onset of supercooling, and the regime close
to the glass transition.

The paper is organized as follows.  We first describe the model and
some technical details in Sec.~\ref{model}.  We then turn to the study
of the relevant timescales, how they are distributed and evolve with
temperature in Sec.~\ref{time}.  We study spatial aspects of the
dynamic in Sec.~\ref{length}: dynamic heterogeneity and four-point
correlations. We compare our numerical results to experimental data in
Sec.~\ref{data}, and conclude the paper in Sec.~\ref{conclusion}.

\section{Model and simulation details}
\label{model}

The NEF model is defined by the Hamiltonian
\begin{equation}
H = \sum_{i=1}^N n_i,
\label{hamiltonian}
\end{equation}
where $n_i = 0,1$ are $N=L^3$ binary variables defined on each site of
a cubic lattice of linear size $L$, which has periodic boundary
conditions.  Physically, $n_i=1$ ($n_i=0$) describes at a
coarse-grained level a site which is mobile (immobile).

The dynamics of the model can be written as follows,
\begin{equation}
n_i=0
\begin{array}{c}
\xrightarrow{~ ~ ~{\cal C}_i ~ c ~ ~ } \\ \xleftarrow[{\cal C}_i ~
(1-c)]{} \\
\end{array}
n_i=1 ,
\label{rates}
\end{equation}
where 
\begin{equation}
c = \langle n_i \rangle =  \frac{1}{1+\exp(1/T)},
\label{concentration}
\end{equation}
is the mean concentration of mobile regions
and ${\cal C}_i$ is the kinetic constraint, 
\begin{equation}
{\cal C}_i = 1-\prod_{\langle j,i \rangle'} (1-n_j) .
\label{kc}
\end{equation}
The prime indicates that the product is restricted to those three
sites, $j$, which are north, front, and east nearest neighbours of
site $i$.  The kinetic constraint (\ref{kc}) implies that site $i$
cannot change state unless at least one of those three neighbours is
in the state $n_j=1$.  The system is therefore more constrained than
the corresponding one-spin facilitated FA model, where any of its six
neighbours can facilitate the site $i$.  In the arrow model,
anisotropy is only local and results physically from the empirical
observation that facilitation has locally a preferred direction that
may persist over a certain length scale even though the system is on
average totally isotropic~\cite{pnas}.  The NEF model corresponds
therefore to having an infinitely large persistence length of dynamic
facilitation. This is physically unrealistic, but constitutes a useful
limit effectively describing systems with a large but finite
persistence of dynamic facilitation \cite{footnote}.

We performed Monte-Carlo simulations of the NEF model for several
temperatures in the range $T \in [0.15, 5.0]$, covering over eleven
orders of magnitude in relaxation times.  Such large time scales can
be simulated using a continuous time algorithm~\cite{Newman-Barkema}.
As we shall see, this dynamical slow-down is accompanied by the growth
of dynamic spatial correlations, and one has to pay attention to
possible finite size effects~\cite{fss}.  This is however not a
serious problem in the present model because even at very large time
scales spatial correlations are not very large (see below).  At the
lowest temperature simulated a system size of $N=32^3$ proved to be
large enough.  Compare this to the system size $N=160^3$ used in the
strong limit to simulate similar timescales~\cite{steve2}.
Equilibrium behaviour is trivial due to the non-interacting
Hamiltonian (\ref{hamiltonian}).  It is straightforward to produce
independent equilibrated initial configurations.  Simulations
therefore only consist of production runs.  Averages are then
performed over truly independent initial configurations.

\section{Time scales}
\label{time}

\subsection{Global dynamics}

\begin{figure}
\psfig{file=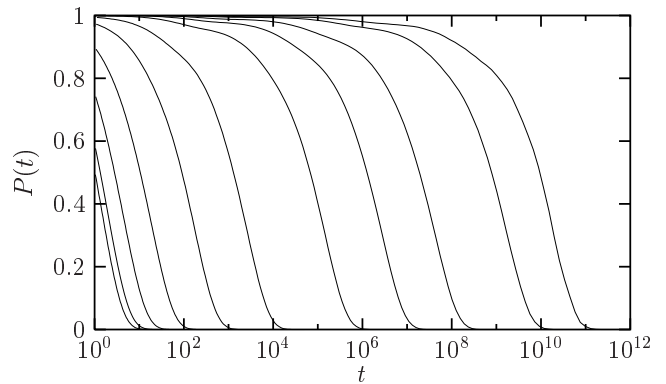,width=8.5cm}
\caption{\label{pers} Persistence function, Eq.~(\ref{perseq}), for
temperatures decreasing from left to right: $T=4.0$, 2.0, 1.0, 0.6,
0.4, 0.3, 0.23, 0.2, 0.18, 0.16, and 0.15.}
\end{figure}

We first consider the spatially averaged dynamics, which 
may be probed via the mean persistence function,
\begin{equation}
P(t) = \left\langle \frac{1}{N} \sum_{i=1}^N
P_i(t) \right\rangle,
\label{perseq}
\end{equation}
where $P_i(t)$ is the single-site persistence function at time $t$,
which takes the value $1$ if site $i$ has not flipped up to time $t$,
and the value $0$ otherwise.  As discussed in
Ref.~\cite{Berthier-et-al}, the persistence function $P(t)$ can be
seen as the analog of the self-intermediate scattering function,
$F_s(k,t)$, of a supercooled liquid at wavelengths comparable to the
particle diameter \cite{footnote2}.  Fig.~\ref{pers} shows, as
expected, that the dynamics slows down markedly when temperature is
decreased below $T_o \approx 1.0$, indicating the onset of slow
dynamics in this model~\cite{pre,Brumer-Reichman}.  Physically, $T_o$
corresponds to the energy scale of the problem, i.e. the energy needed
to create an excitation, see Eq.~(\ref{concentration}).

The shape of the mean persistence function will be discussed below in
some detail. For the moment, we simply note that the long-time decay
is non-exponential, as is commonly observed in supercooled
liquids. Moreover, the short-time dynamics also presents non-trivial
features, see for instance the curve for $T=0.15$ in Fig.~\ref{pers}.

\begin{figure}
\psfig{file=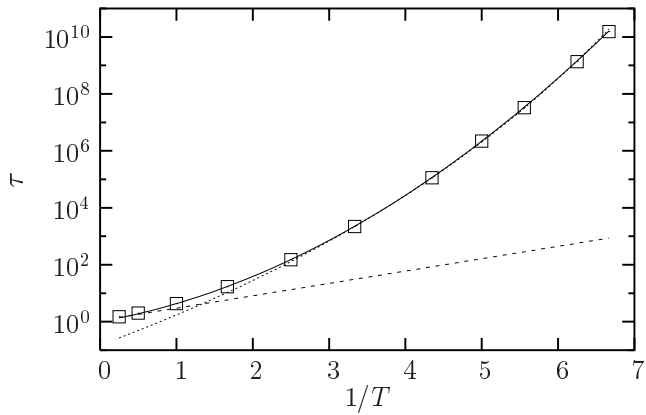,width=8.5cm}
\caption{\label{tauT} Points: Mean relaxation time from $P(\tau) =
e^{-1}$. The lines represent various fits.  Dashed line is the
high-temperature Arrhenius behaviour, $\tau \sim \exp(1/T)$. Full line
is $\tau(T) \sim \exp(\Delta(T)/T)$ with $\Delta(T) = 0.634/T+1.1$.
The dotted line shows that a Vogel-Fulcher form, $\tau \sim
\exp[2.3/(T-0.06)]$, fits the low temperature data with an unphysical
finite $T$ singularity.}
\end{figure}

The mean relaxation time, $\tau(T)$, is extracted via the usual
relation $P(\tau) = e^{-1}$.  In the present context, $\tau$
represents also the $\alpha$-relaxation of the model.  The temperature
dependence of $\tau$ is shown in Fig.~\ref{tauT}.  The main
observation from this figure is that the mean-relaxation time grows
with decreasing temperature in a super-Arrhenius manner, see
Fig.~\ref{tauT}. The NEF model behaves as a fragile supercooled
liquid, as expected \cite{pnas}. Note however that 
the ratio $T_o / T_g \approx 7$ that can be extracted 
from Fig.~\ref{tauT} is about three times smaller 
in fragile liquids such as orthoterphenyl. 

Various fits are also included in Fig.~\ref{tauT}.  The high
temperature behaviour is well described by a naive mean-field
approximation~\cite{eisinger,pre},
\begin{equation} 
\tau_{MF} \sim c^{-1}.
\end{equation}
This behaviour breaks down below $T_o$, where dynamics is dominated by
local fluctuations of the mobility.  Generalizing to three dimensions
the results for the East model~\cite{east,Aldous-Diaconis,pnas}, we
expect that in the limit of small temperature, the time scale behaves
as
\begin{equation}
\tau \sim c^{-\Delta(T)},
\label{bassler}
\end{equation}
with $\Delta(T) \approx  b/T$ at low $T$, where 
$b$ is a constant. Since $c \sim e^{-1/T}$,
Eq.~(\ref{bassler}) implies an exponential inverse temperature
squared dependence at low temperatures, $\tau \sim e^{b/T^2}$.
In fact, an empirical fit using
\begin{equation}
\Delta(T) = a + \frac{b}{T},
\label{bassler2}
\end{equation}
allows one to describe the whole temperature range studied, using
$b=0.634$ and $a=1.1$.  The supposedly exact low-temperature result,
$\Delta(T) = (3 \ln 2 T)^{-1}$, works well for very large relaxation
times. Indeed we find that a plot of $T^2 \log \tau$ vs. $1/T$
nicely converges to the theoretical value $(3 \ln 2)^{-1}$ (not shown).
At low temperatures, the $\alpha$-relaxation in the NEF model
therefore follows a B\"assler law~\cite{bassler}.

Fig.~\ref{tauT} also shows that the low $T$ data can be fit using the
Vogel-Fulcher law, $\tau_{VF} \sim \exp \left[ A/(T-T_0) \right]$.
The B\"assler and Vogel-Fulcher curves are hardly distinguishable for
over eight orders of magnitude, as has been observed before when
fitting experimental data~\cite{bassler}.  In our case, it is evident
that the finite temperature singularity of the Vogel-Fulcher law is
completely unphysical.

As will be discussed below, the fragile behaviour of the NEF model
results from the existence of a hierarchy of lengthscales governing
the dynamical behaviour.  It is therefore not surprising that the
relaxation time in this model does not follow the Adam-Gibbs relation
which is argued to link dynamics and thermodynamics of supercooled
liquids~\cite{ag,angell}.  For the NEF model, the entropy has
essentially an Arrhenius behaviour at low temperatures.  The
Adam-Gibbs relation thus predicts the relaxation time to grow as a
double exponential of the inverse temperature.  This vastly
overestimates the growth of $\tau$, which follows the B\"assler law,
Eqs.\ (\ref{bassler},\ref{bassler2}).  In the same vein, naively
extracting a Kauzmann temperature by linearly extrapolating the
entropy to zero in the same temperature range where the Vogel-Fulcher
law seems to apply (see Fig.\ \ref{tauT}), gives a value $T_K \approx
0.16$, which is very different from the Vogel-Fulcher temperature $T_0
\approx 0.06$ found above.  Although $T_0$ and $T_K$ are ill-defined
temperatures in our case, experimentally they are often found to be
close in fragile liquids, a feature which is not quantitatively
reproduced by the NEF model.

\subsection{Distribution of relaxation times}

\begin{figure}
\psfig{file=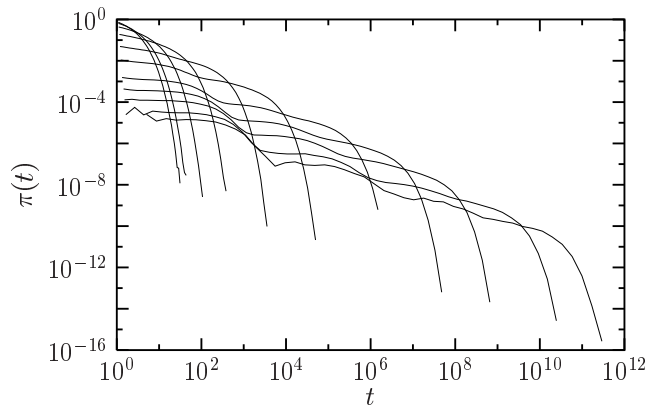,width=8.5cm}
\caption{\label{dis} Distribution of persistence time $\pi(t)$, see
Eq.~(\ref{pi}), for temperatures as in Fig.~\ref{pers}, decreasing
from left to right.}
\end{figure}

The mean relaxation time $\tau(T)$ captures only in part the
relaxation behaviour of the model.  We consider in this subsection the
distribution of relaxation times, $\pi(t)$, related to the mean
persistence function via~\cite{pre}
\begin{equation}
P(t) = \int_t^\infty dt' \pi(t').
\label{pi}
\end{equation}
These distributions are shown in Fig.~\ref{dis}.

As mentioned above, the long-time tail of the persistence function is
well-described by a stretched exponential form, which implies the
following long-time behaviour for the distribution of relaxation
times,
\begin{equation}
\pi(t) \sim \frac{t^{\beta-1}}{\tau^\beta} \exp
\left[ - \left( \frac{t}{\tau} \right)^\beta \right],
\label{stretched}
\end{equation}
valid for when $t \gg \tau$.  As for the East
model~\cite{Sollich-Evans,Buhot,jcp}, we find that the stretching
exponent $\beta$ decreases with $T$ in the regime $T<T_o$, from its
high temperature value $\beta = 1$.

\begin{figure}
\psfig{file=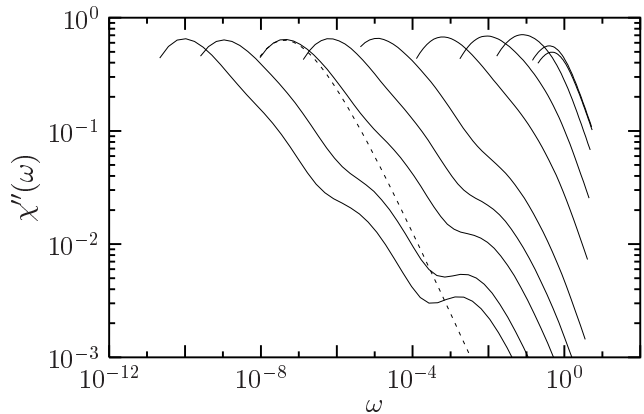,width=8.5cm}
\caption{\label{wing} Imaginary part of the dynamic susceptibility
defined in Eq.~(\ref{susceptibility}), for temperatures as in Fig.~\ref{pers},
decreasing from right to left.  For $T=0.18$, we also show as a dashed
line the Fourier transform of the stretched exponential fit to the
long time tail of the persistence function, revealing `additional
processes' on the `high-frequency flank' of the $\alpha$-relaxation.}
\end{figure}

As can be guessed from Figs.~\ref{pers} and \ref{dis}, using $\tau(T)$
and $\beta(T)$ as unique fitting parameters does not allow for a
satisfactory description of the whole decay of the persistence
function and distribution of relaxation times.  This becomes evident
when data are presented in an alternative way.  Following the
experimental literature, we show the data obtained from the
distribution of time scales in a frequency representation
via~\cite{bloch}
\begin{equation}
\chi''(\omega) = {\rm Im}~\int_{-\infty}^\infty \pi(\log(\tau))
 \frac{1}{1+i \omega \tau} \, d \log \tau ,
\label{susceptibility}
\end{equation}
as is often done with dielectric susceptibility measurements.  This
will also allow us to make quantitative comparisons to experiments
below.  The results for $\chi''(\omega)$ are displayed in
Fig.~\ref{wing}.  At temperature $T=0.18$ we show on top of the data
the Fourier transform of the stretched exponential fit to the long
time tail of $P(t)$. The fit is clearly incorrect in this
representation, although it would look acceptable in most of a plot
such as in Fig.~\ref{pers}.  In fact, deviations appear when $\omega
\tau > 1$, corresponding to $t < \tau$ in the time domain.  This
reveals the presence of `additional processes' on the `high-frequency
flank' of the $\alpha$-relaxation.

This feature is obviously reminiscent of the `high-frequency', or
`Nagel' wing extensively studied by dielectric spectroscopy in several
materials~\cite{nagel}.  The wing is usually observed in fragile
glass-formers, but its precise nature has not been fully elucidated.
Despite initial claims of certain universal behaviour of the
high-frequency wing, more recent investigations seem to favour the
interpretation that the phenomenon does not obey universal 
scaling~\cite{nagel}.
This is compounded by the observation that the amplitude of the
phenomenon also seems to depend on the technique used to study
it~\cite{rossler-epjb}.

In our case, the physical interpretation of the wing is very clear.
The relaxation of the NEF model proceeds in a hierarchical manner,
just as in the one-dimensional
case~\cite{garrahan-chandler,Sollich-Evans}.  In order to relax a
domain of immobile regions of size $\ell$, a region of size $\ell/2$
must first be relaxed, which itself necessitates the relaxation of a
domain of size $\ell/4$, etc., down to the smallest size
$\ell=1$. This implies an energy cost $\Delta E (\ell) \sim \ln \ell$,
from which the B\"assler form, Eq.(\ref{bassler2}), and stretched
exponential decay, Eq.~(\ref{stretched}),
follow~\cite{garrahan-chandler}.  Interestingly, the hierarchy also
shows up in the distributions of Figs.~\ref{dis} and \ref{wing} at
time scales shorter than the $\alpha$-relaxation in a manner
reminiscent of the experimental finding of additional short-time
processes.  Due to the underlying lattice structure of the model, the
hierarchy is discrete rather than continuous, as can be indeed
observed in Fig.~\ref{wing}.  Notice also that there is no `fast'
process at the microscopic time scale, $\omega \sim 1$, in our
system. This is a consequence of coarse-graining by which molecular
vibrations are removed.

These results show that the pattern of dynamical relaxation of the NEF
model is quantitatively accurate on a wide frequency range. This will
be shown explicitly in Sec.~\ref{data} where dielectric susceptibility
data taken close to the experimental glass transition are compared to
the NEF model results.  We recall that in the strong case the
large-time decay was found to be purely exponential, while short-time
processes had a much smaller magnitude~\cite{steve2}.  This prediction
is however hard to confirm experimentally since strong liquids are not
easy to study by dielectric spectroscopy~\cite{leheny}.

Physically, our results also suggest that the high-frequency wing
observed in fragile glass-formers is an intrinsic feature of the
$\alpha$-relaxation linked in a direct way to their dynamical
hierarchical structure which is also at the origin of stretched
relaxation and fragile behaviour.

When the relaxation time is moderate, an even more complex behaviour
is observed due to the influence of clusters of defects. Such dynamic
objects are irrelevant at low temperatures, but can influence the
short-time dynamics just below $T_o$, as discussed in detail in
Ref.~\cite{pre}.  These clusters were in particular shown to be
responsible for the temperature behaviour of a number of quantities
discussed in some numerical works.  Clusters also produce additional
short-time processes visible in the distributions $\pi(t)$ when
temperature is not too low~\cite{pre}.  Interestingly in the present
context, these patterns closely resemble the ones observed
experimentally in mildly supercooled liquids and usually explained in
terms of mode-coupling theory~\cite{mct}. This will be demonstrated in
Sec.~\ref{data}, where data recently fitted via mode-coupling
theory~\cite{sperl} are also fitted using the NEF model.

\section{Length scales}
\label{length}

\subsection{Dynamic heterogeneity}

\begin{figure}
\psfig{file=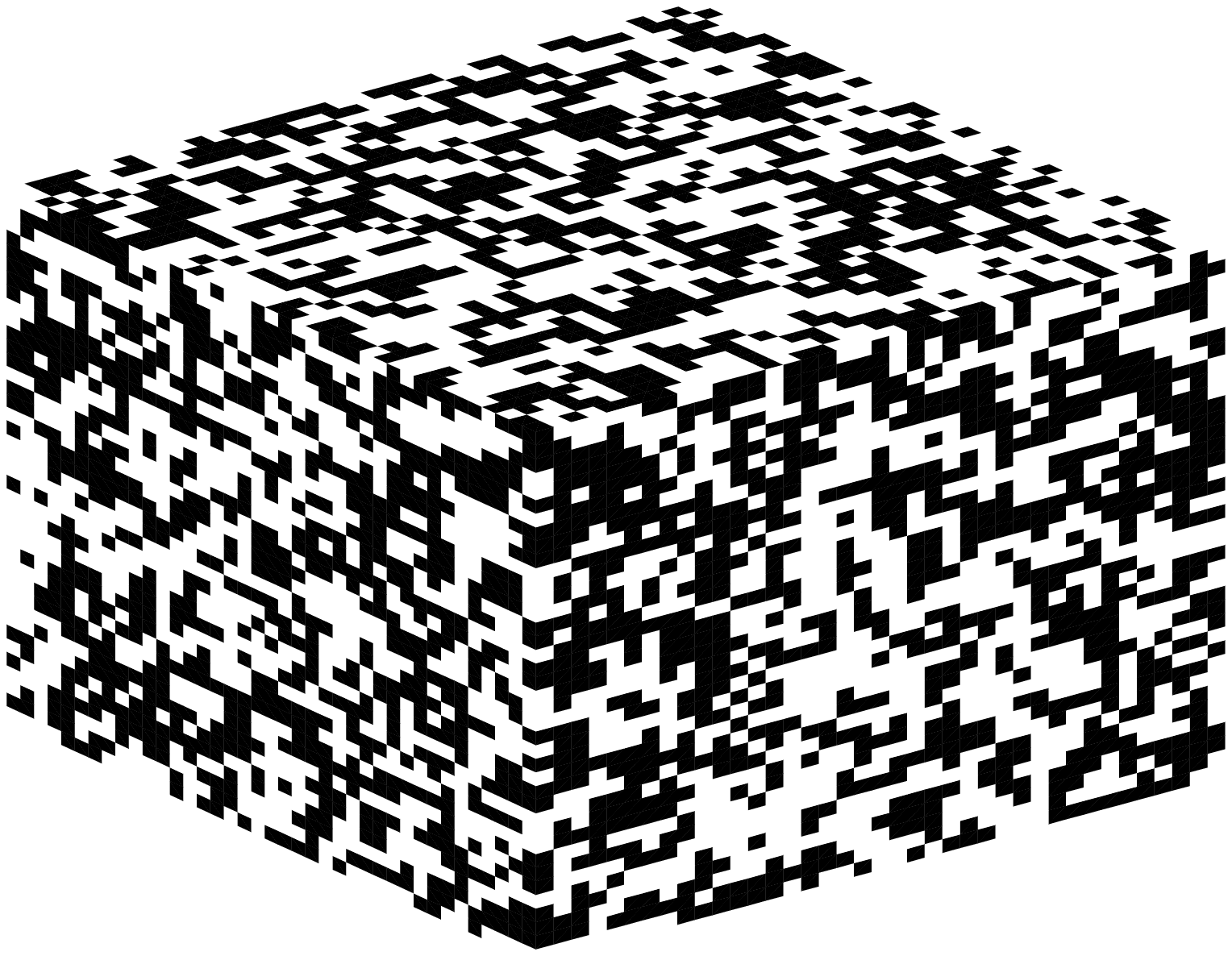,width=6.cm,height=6.cm}
\psfig{file=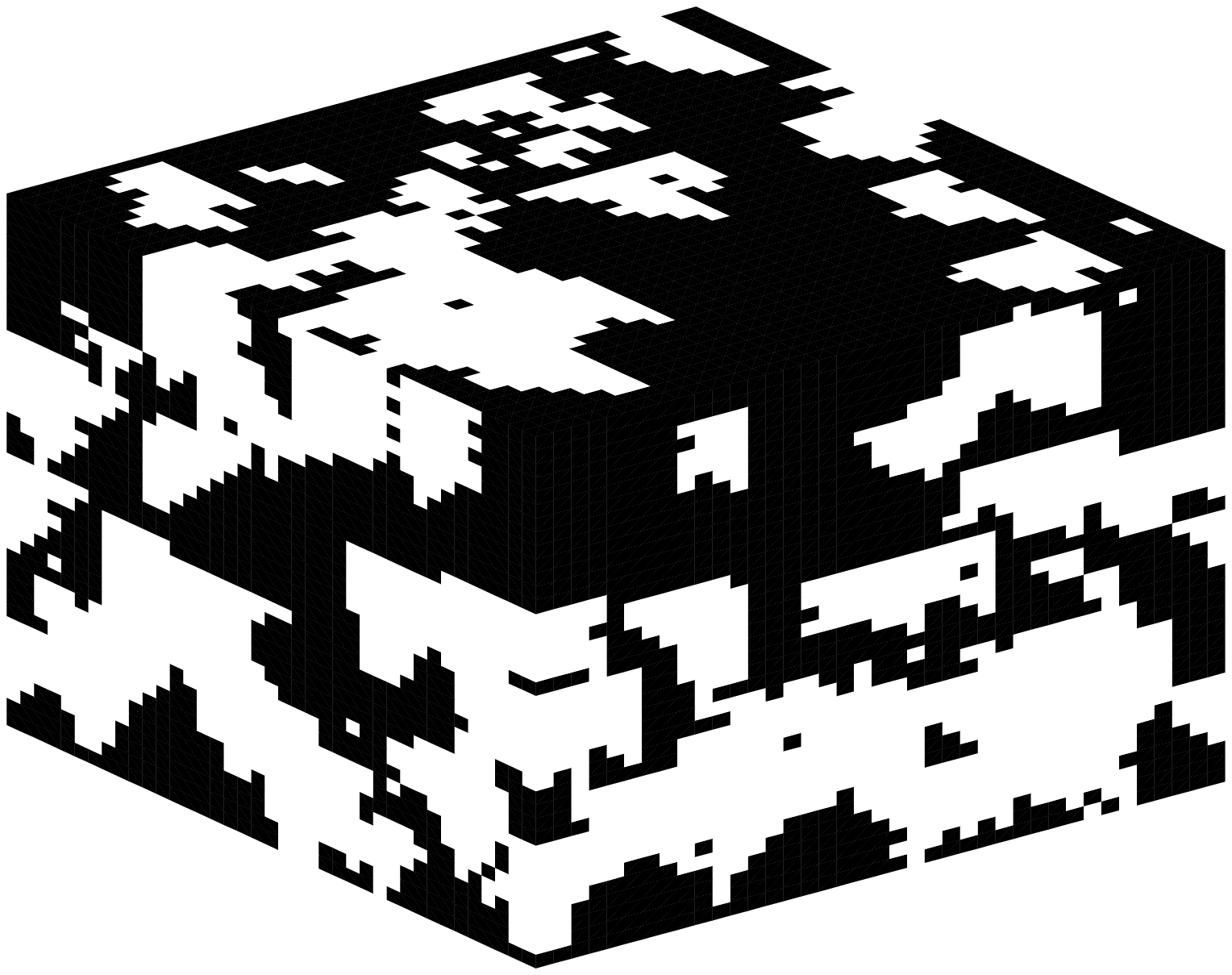,width=6.cm,height=6.cm}
\caption{\label{critical} Spatial distribution of the local
persistence at time $t^\star$ such that $P(t^\star)=1/2$ (i.e., 50\%
of sites, shown in black, have flipped by time $t^\star$) at
temperatures $T=1.0$ (top) and $T=0.15$ (bottom) for a system size
$L=40$ in both cases.  The appearance of dynamic critical fluctuations
when $T \to 0$ is evident.}
\end{figure}

\begin{figure}
\psfig{file=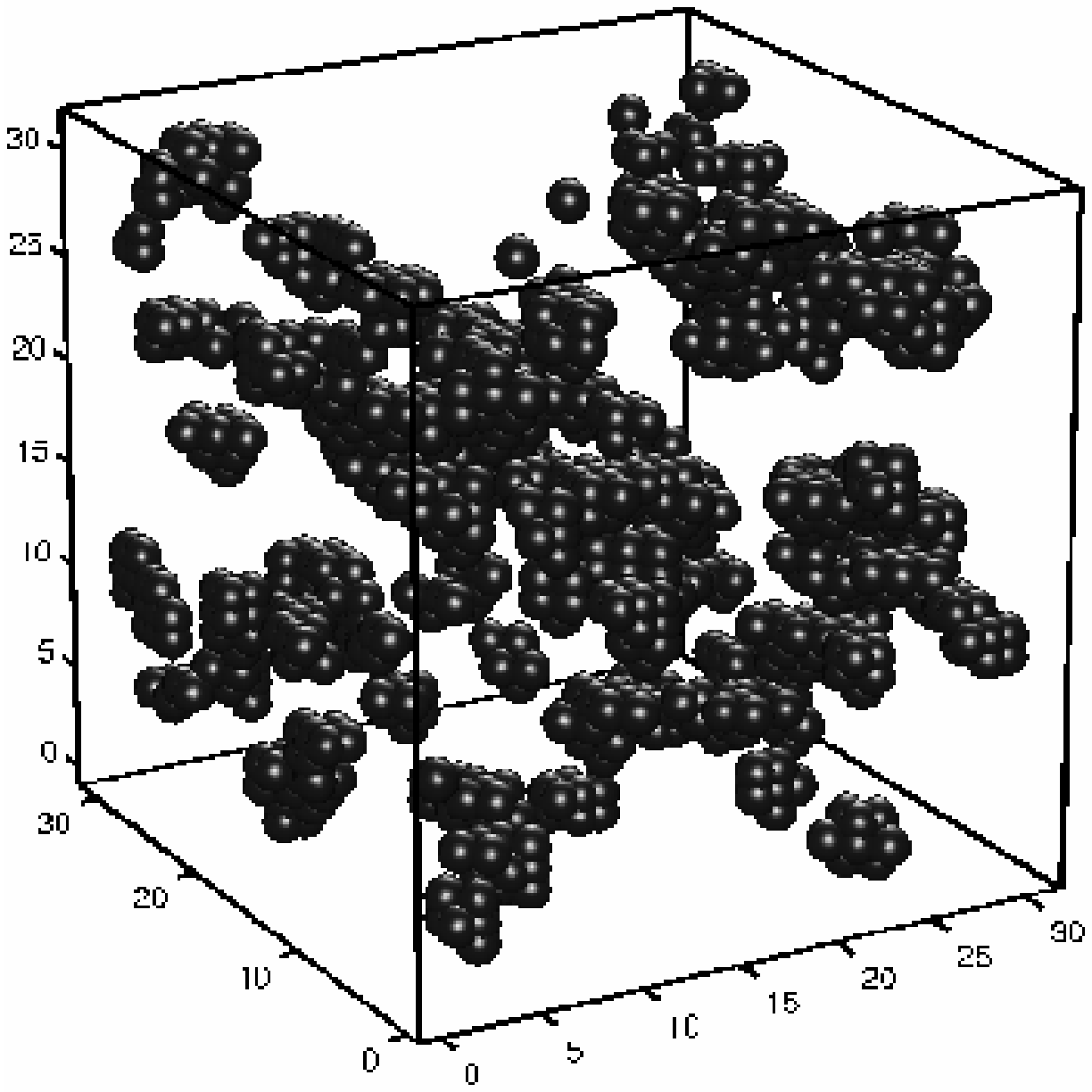,width=6cm}
\psfig{file=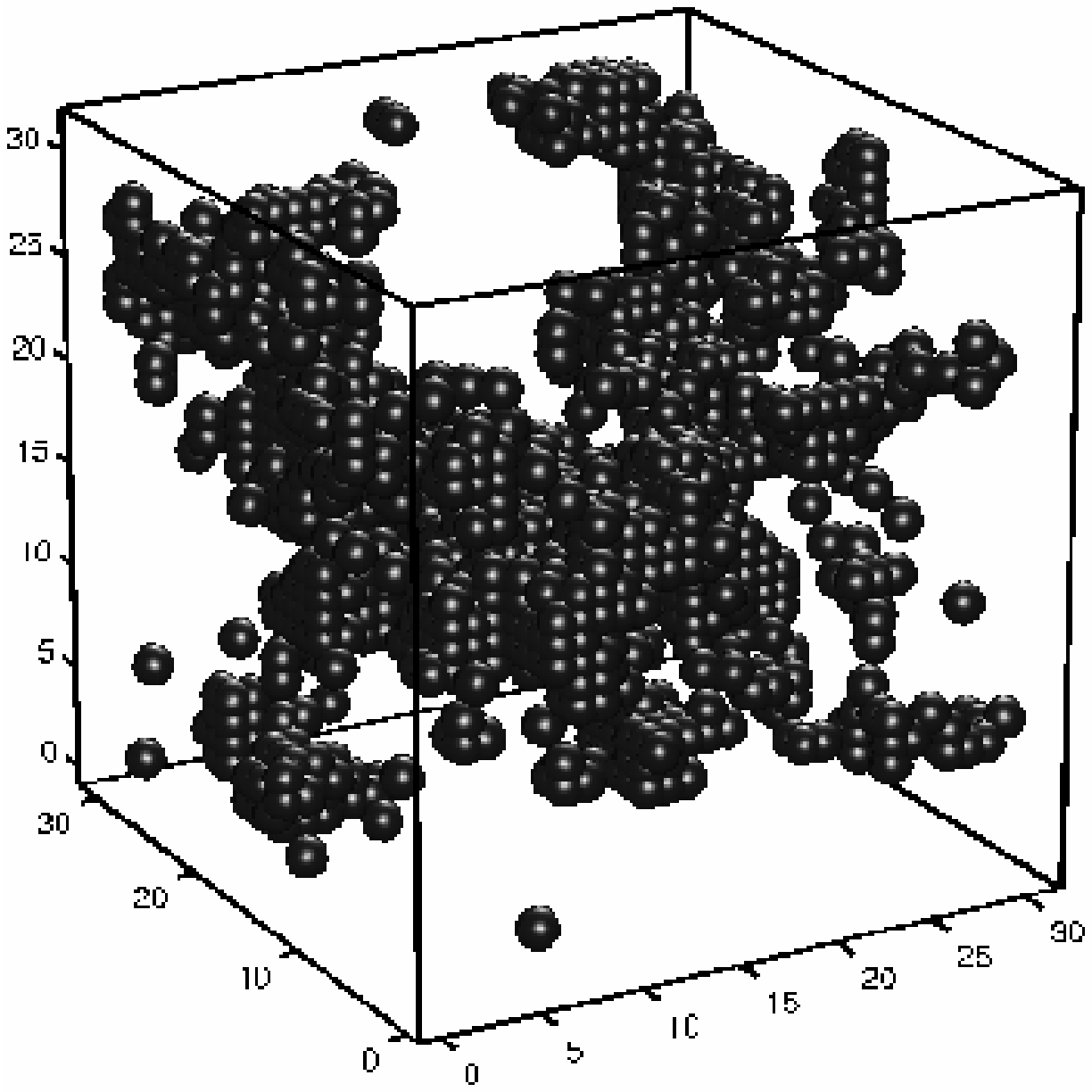,width=6cm}
\caption{\label{bubble} Visualization of 5\% of the sites that relax
faster (top) and slower (bottom) in a given run at $T=0.18$ for a
system size $L=30$.  Fast/slow sites are painted as spheres to
highlight the similarity between heterogeneity in lattice models to
that in atomistic models.  The fast clusters in the top figure are
less compact than the slow ones in the bottom figure as a 
consequence of relaxation via 
point like excitations. See the text for details.}
\end{figure}

The growth of timescales in the NEF model is accompanied by growing
spatial correlations as the system approaches its critical point at
$T=0$.  These correlations are purely dynamical in origin, and give
rise to dynamic heterogeneity~\cite{garrahan-chandler,Berthier-et-al}.
Figures~\ref{critical} and \ref{bubble} serve to
illustrate this phenomenon, as we now explain.

We quantify the local dynamics via the local persistence function
$P_i(t)$.  For a given temperature we run the dynamics for a time
$t^\star$, such that $P(t^\star) = 1/2$, meaning that half of the
sites have flipped at least once. We colour white persistent
(immobile) spins, for which $P_i(t^\star)=1$, and black transient
(currently or previously mobile) spins, for which $P_i(t^\star) =
0$. Figure~\ref{critical} shows the local persistence function for the
NEF model at two different temperatures, $T_o=1.0$ and $T=0.15 \ll
T_o$.  Clearly, the low-temperature dynamics is spatially
heterogeneous, and the spatial correlations of the local dynamics grow
as $T$ is decreased. The `critical' nature of dynamic clusters is
apparent: the pictures are reminiscent of the spatial fluctuations of
an order parameter close to a continuous phase transition, such as the
magnetization of an Ising model near criticality. In our case, the
order parameter is a dynamic object, the persistence function, and the
critical fluctuations are purely dynamical in origin~\cite{fss}.

It is interesting to note that these figures are qualitatively
different from the ones obtained in the strong case where dynamic
facilitation is isotropic.  One can clearly distinguish in
Fig.~\ref{critical} the North, East and Front directions of
facilitation, implying that wandering of excitations in the other
three directions is forbidden.  Domains of Fig.~\ref{critical} appear
much less rough than the ones obtained in the isotropic case
\cite{steve2}.  In that sense, increasing the fragility is similar to
increasing the `surface tension' of the dynamic domains observed in
Fig.~\ref{critical}.  The same observation applies to an even more
fragile system, the two-spin facilitated FA model in two dimensions,
where the corresponding domains resemble a polydisperse assembly of
squares~\cite{harrowell}.

The qualitative observations of dynamic heterogeneity performed in
numerical or experimental works can also be made in the present
coarse-grained model.  We show in Fig.~\ref{bubble} the analog of
spatial clustering of subsets of fast and slow sites. To build these
snapshots, we represent a given percentage, 5\%, of the sites that
relax faster or slower, i.e. we show those sites for which the
persistence time is among the 5\% smaller or larger in a randomly
chosen run at temperature $T=0.18$.

One observes that the fastest sites are not randomly located in space,
but clustered in `non-compact' or `stringy' objects, similar to those
observed in simulations and
experiments~\cite{string,harrowell2,weeks}.  The shape of these
objects is a consequence of the existence of point defects of
mobility. When a defect moves, it induces those sites along its
trajectory to relax, so leaving in its wake a string of fast sites.

In Fig.~\ref{bubble} we also show 5\% of the sites which are slowest,
using the same simulation temperature as before.  A more compact
structure is seen. This is again the consequence of the relaxation via
point defects of mobility. The slowest sites belong to regions of
space devoid of defects which take then a very long time to be visited
by defects.  These large domains are thus slowly relaxed.  It is the
bulk of these slow domains that is observed in Fig.~\ref{bubble}.
Note finally that at large times, the distribution of slow cluster sizes
seems very wide, since some isolated sites which have been not been
visited by defects coexist with the very large domains 
discussed above. 

Since our comments on Fig.~\ref{bubble} are mainly qualitative, it 
should come as no surprise that snapshots built in this fashion 
in the isotropic case are very similar~\cite{steve2}.

Finally, it is interesting to compare these figures to those in previous
publications~\cite{garrahan-chandler, pre}, in which space-time
diagrams of one dimensional kinetically constrained models were
presented.  There, spatio-temporal `bubbles' of immobile regions,
bounded by diffusing point defects were presented.  In the present
model these bubbles become $(3+1)$-dimensional objects, and
Fig.~\ref{bubble} their three-dimensional 
spatial projections of trajectories of
various time extensions.

\subsection{Spatial correlations}

We now quantify these qualitative
observations via appropriate statistical
correlators.  We measure spatial correlations of the local dynamics
via the following spatial correlator of the local persistence
function,
\begin{equation}
\label{c}
C(r,t,T) = \frac{1}{N f(t)}\sum_{i=1}^N 
\Big[ \langle P_i(t) P_{i+r}(t)
\rangle - P^2(t) \Big],
\end{equation}
where the function $f(t) = P(t)-P^2(t)$ in the denominator ensures the
normalization $C(r=0,t,T)=1$.  Alternatively, one can take the Fourier
transform of (\ref{c}), giving the corresponding structure factor of
the dynamic heterogeneity,
\begin{equation}
\nonumber 
S(q,t,T) = \frac{1}{N f(t)} \sum_{k,l=1}^N \Big[ \langle P_k(t)
P_l(t) \rangle - P^2(t) \Big] e^{i q ( k - l )},
\end{equation}
for wavevectors defined in the first Brillouin zone of the cubic lattice.

\begin{figure}
\psfig{file=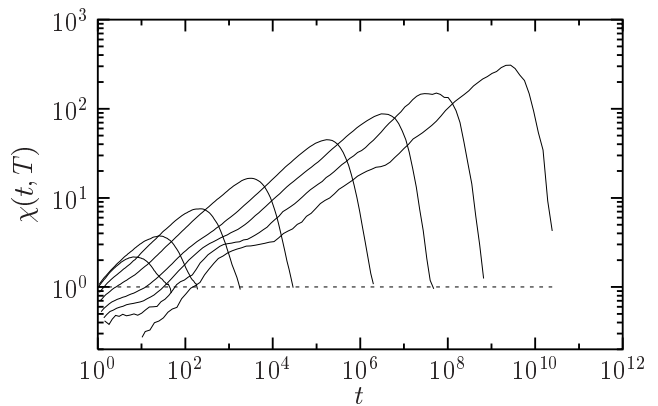,width=8.5cm}
\caption{\label{chi4} Dynamic `four-point' susceptibility,
Eq.~(\ref{chichi}), at temperatures $T=1.0$, 
0.6, 0.4, 0.3, 0.23, 0.2, 0.18, and 0.16 (from left to right).}
\end{figure}

Finally, the zero wavevector limit of $S(q,t,T)$ defines a dynamic
`four-point' susceptibility, $\chi(t,T) = S(q=0,t,T)$, which can be
rewritten as the normalized variance of the (unaveraged) persistence
function, $p(t) \equiv N^{-1} \sum_{i=1}^N P_i(t)$,
\begin{equation}
\label{chichi}
\chi(t,T) = \frac{N}{f(t)} \big[ \langle p^2(t) \rangle - \langle p(t)
\rangle^2 \big].
\end{equation}
It should be noted that normalizations also ensure that $\chi(t,T)$
remains finite in the thermodynamic limit, except at a dynamic
critical point such as the ones discussed in Refs.~\cite{steve1,mct2}.

Figure~\ref{chi4} shows the time dependence of the susceptibility
(\ref{chichi}) for various temperatures. The behaviour of $\chi$ is
similar to that observed in atomistic simulations of supercooled
liquids~\cite{DHreviews3}.  The susceptibility develops at low
temperature a peak whose amplitude increases, and whose position
shifts to larger times as $T$ decreases.  As expected, the location of
the peak scales with the relaxation time $\tau(T)$, indicating that
dynamical trajectories are maximally heterogeneous when $t \approx
\tau(T)$.

\begin{figure}
\psfig{file=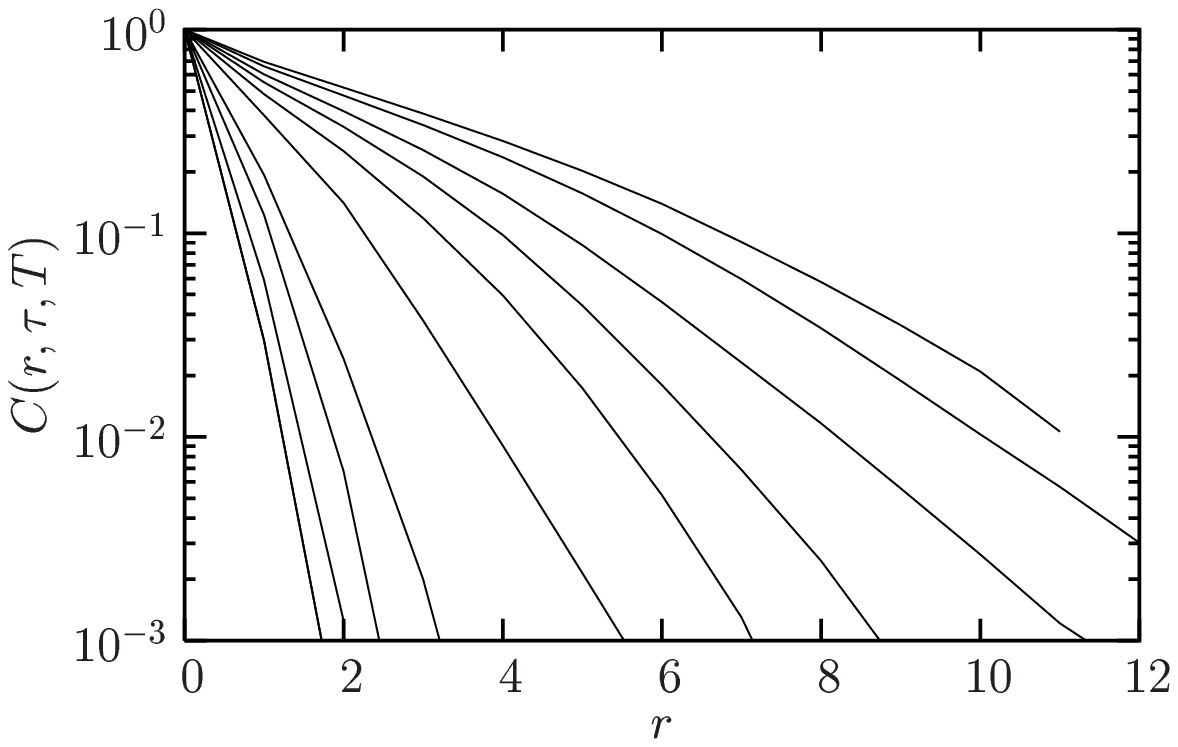,width=8.5cm}
\psfig{file=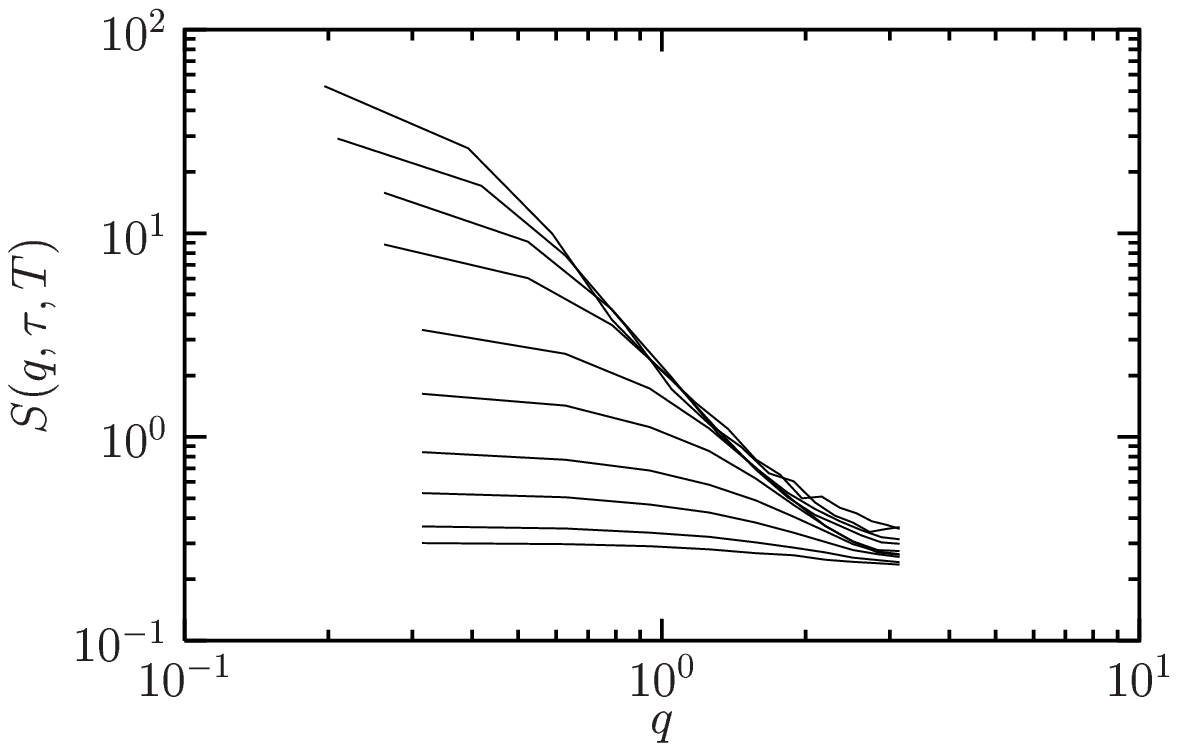,width=8.5cm}
\caption{\label{cr} Top: Spatial correlator of dynamic heterogeneity 
at the relaxation time $\tau(T)$.
Temperatures as in Fig.~\ref{chi4} 
decrease from left to right.
Bottom: corresponding structure factor.
Temperatures as in Fig.~\ref{chi4} 
decrease from bottom to top.}
\end{figure}

In Figure~\ref{cr} we show the correlator $C(r,t,T)$ and the structure
factor $S(q,t,T)$ for different temperatures and fixed times
$t=\tau(T)$ where dynamic heterogeneity is maximal.  These correlation
functions confirm, as suggested by Fig.~\ref{critical}, that a dynamic
length scale associated with spatial correlations of mobility develops
and grows as $T$ decreases: spatial correlations decay more slowly
with distance as $T$ decreases, while a peak in the structure develops
and grows at $q=0$.

\begin{figure}
\psfig{file=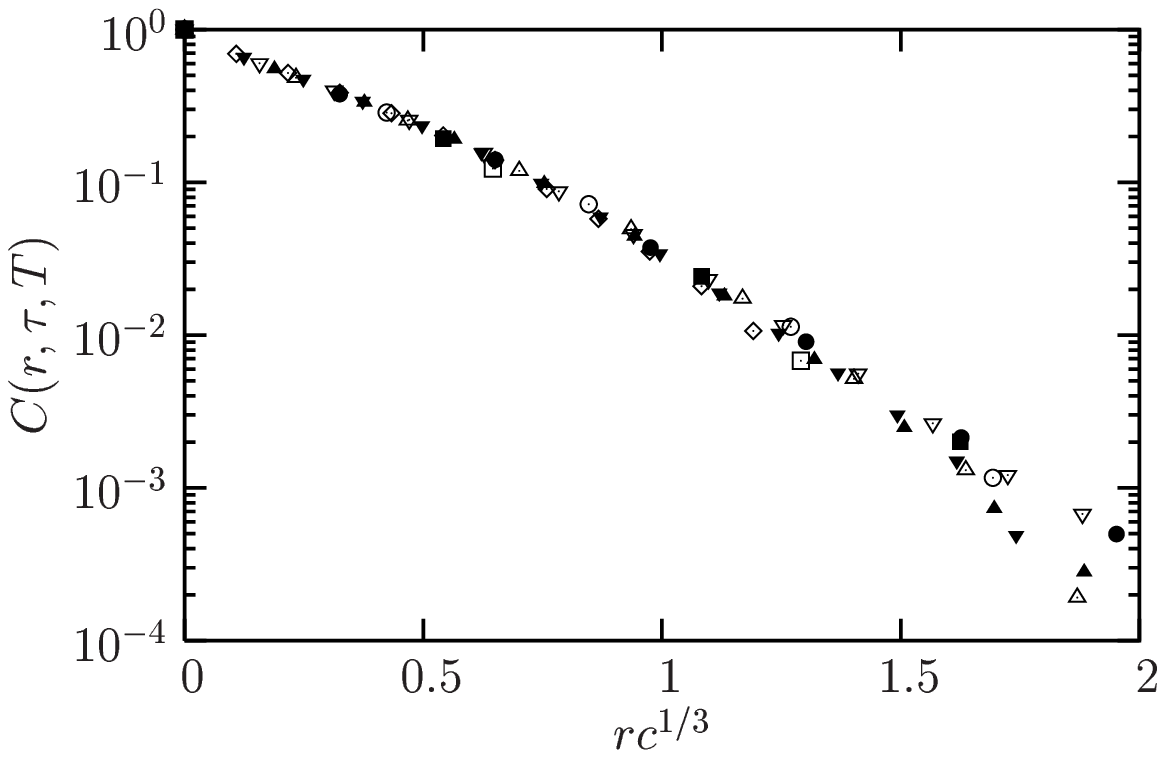,width=8.5cm}
\psfig{file=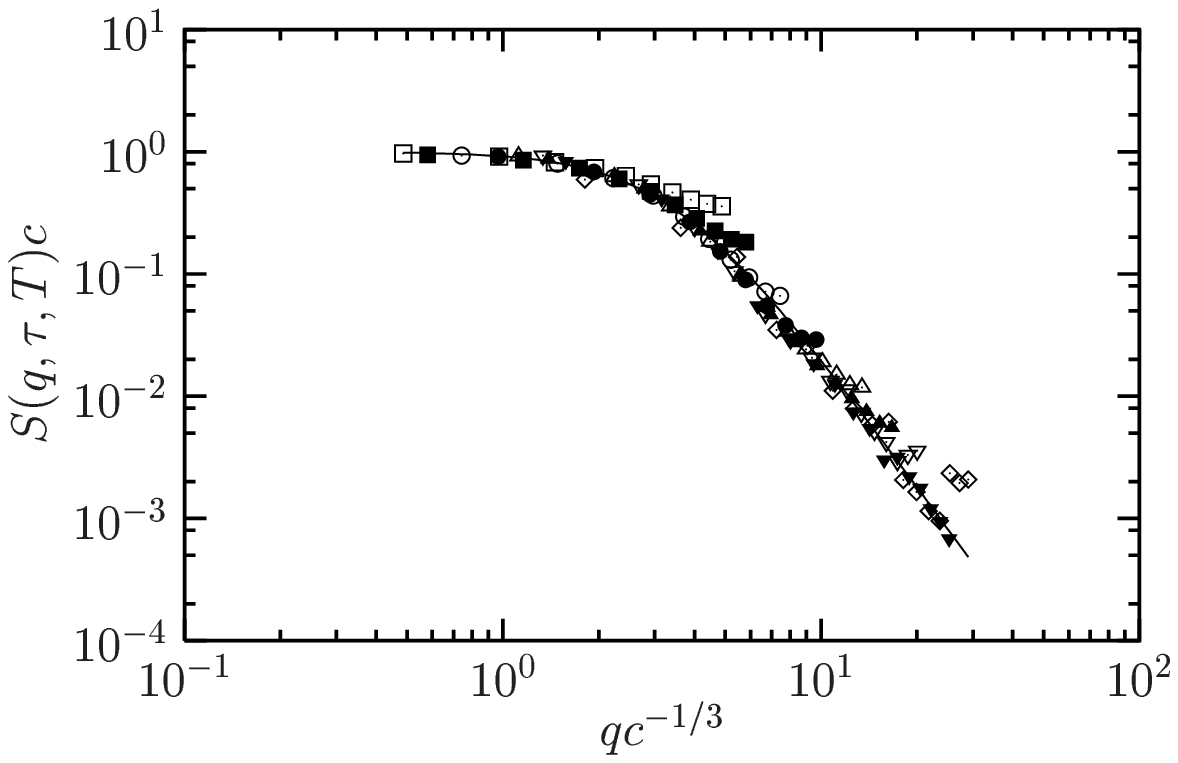,width=8.5cm}
\caption{\label{scal} Top: Spatial correlator rescaled with the form
$C(r) \sim {\cal C}(r/\xi)$, $\xi \sim c^{-1/3}$.  Bottom: Rescaled
structure factor with $\chi = S(q=0) \sim c^{-1}$ and $\xi \sim
c^{-1/3}$. Full line interpolates between $S(q\to 0) \sim const$ and
$S (q \to \infty) \sim k^{-3.58}$.  }
\end{figure}

It is possible to extract numerically the value of the corresponding
dynamic length scale, $\xi(T)$, at each temperature. To do so, we
study in detail the shape of the correlation functions shown in
Fig.~\ref{cr}. As for standard critical phenomena, we find that the
dynamic structure factor can be rescaled according to
\begin{equation}
S(q,\tau,T) \sim \chi(\tau,T) \, {\cal S} \left( q \xi \right),
\label{scalnum}
\end{equation}
where the scaling function ${\cal S}(x)$ has the following 
asymptotic behaviours in the scaling regime, $\xi \gg 1$:
\begin{eqnarray}
{\cal S}(x \to 0) \sim & const, \\
{\cal S}(x \to \infty) \sim & x^{2-\eta}.
\label{asympt}
\end{eqnarray}
Both the susceptibility $\chi$ and the dynamic length scale $\xi$
estimated at time $t=\tau$ behave as power laws of the defect
concentration,
\begin{equation}
\chi \sim c^{-\gamma}, \quad \quad \xi \sim c^{-\nu}.
\label{exp}
\end{equation}
These relations imply that the exponents $\gamma$ and $\nu$ should be
numerically accessible by adjusting their values so that a plot of
$c^{\gamma} S$ versus $q c^{-\nu}$ is independent of temperature. We
show such a plot in Fig.~\ref{scal}, and we find that the values
$\gamma \approx 1$ and $\nu \approx 1/3$ lead to a good collapse of
the data.  The exponent $\gamma$ can be independently and more
directly estimated from Fig.~\ref{chi4} by measuring the height of the
maximum of the susceptibility for various concentrations. The result
is also well fitted to the power law $c^{\gamma}$ with $\gamma \approx
1$.  We find also that the scaling function $S(x)$ is well-described
by an empirical form $S(x) = 1/(1+x^{2-\eta})$, consistent with
Eq.~(\ref{asympt}). Thus we can determine the value of the `anomalous'
exponent, $\eta$.  We find that the theoretically expected $\eta =
-\ln 3 / \ln 2 \approx -1.58$ describes the data very
well~\cite{garrahan-chandler}.  The fact that this exponent is much
more negative than in the strong case of the FA model quantifies our
observation that domains of Fig.~\ref{critical} were much less rough
in the NEF model than in the FA model, i.e., that dynamic domains of
fragile systems have smoother boundaries than those of strong ones.

The spatial correlator $C(r,\tau,T)$ also obeys scale invariance, and
we find that
\begin{equation}
C(r,\tau,T) \sim {\cal C} \left( \frac{r}{\xi},\tau \right),
\label{crscal}
\end{equation}
with the asymptotic behaviour ${\cal C}(x \to 0) \sim x^{-0.58}$, as a
result of a generalized Porod's law.  The scaling behaviour
(\ref{crscal}) is presented in Fig.~\ref{scal}, which also confirms
the temperature dependence of the correlation length, $\xi \sim
c^{-1/3}$ \cite{pnas}.
It is interesting to note that the 
relation between susceptibility and length is very different 
in the strong and fragile cases, since we find that 
$\chi \sim \xi^3$ in the NEF model, while 
the scaling is closer to $\chi \sim \xi^2$
in the FA case~\cite{steve2}.

\section{Comparison to experimental data}
\label{data}

In this section, we use the numerical results obtained in previous
sections to fit experimental data with the NEF model.  In doing so
there are several points that need to be considered.

First, the model we use is meant to be a description of supercooled
liquids which is coarse-grained both in time and space, and lives on a
lattice.  We are thus dealing with discrete, rather than continuous,
spatial degrees of freedom, and the very short-time dynamics of the
liquid is removed.  By construction, this produces discrepancies
between real data on liquids and numerical NEF model data, especially
for short times and small lengths.  This is a small price to pay in
such an approach given the large number of features that can still be
satisfactorily accounted for with the NEF model.

Second, we have some freedom on how to relate real experimental
timescales and Monte Carlo steps in the simulation. This will be done
empirically, and we find that the expected equivalence, 1 MC step
$\approx$ few ps~\cite{pnas}, works well, independently of the
temperature. Explicitly, we found that 1 MC $\in [1,10]$ ps
for the whole range of experimental data we have 
considered~\cite{nagel,fayer}.

Third, one must adjust the temperature given in Kelvin in experiments
to the adimensional $T$ of the simulations. This amounts to fitting
the value of an energy scale, $J$, which should appear on dimensional
grounds in front of the Hamiltonian (\ref{hamiltonian}).  In
principle, $J$ could be fixed independently by fitting, say, viscosity
data of a given liquid before using the corresponding temperatures to
fit more detailed dynamic data in what would be a zero-parameter
fitting procedure.  We have done the correspondence in a less
constrained manner, adjusting $T$ in the simulation to give a good fit
to the data.  Very satisfactorily, though, we end up with a
correspondence between numerical and experimental temperatures which
is well described by linear relations, as it should be at low
temperatures \cite{pnas}.

\begin{figure}
\psfig{file=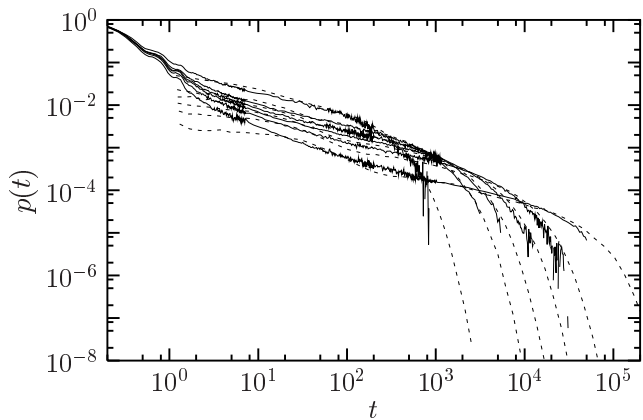,width=8.5cm}
\caption{\label{bpm} Optical Kerr Effect experimental for supercooled
liquid 2-biphenylmethanol (full lines from Ref.~\cite{fayer}) at
temperatures $T=359$ K, 327 K, 319 K, 311 K, 303 K, 291 K (from left
to right).  Dashed lines are corresponding data for the NEF model at
temperatures $T=0.4$, 0.34, 0.32, 0.3, 0.28 and 0.25 (from left to
right).  Time is counted in picoseconds in the experiments, in Monte
Carlo steps in simulations.}
\end{figure}

In Fig.~\ref{bpm}, we show the result of this fitting procedure as
applied to recent data measured using the so-called optical Kerr
effect~\cite{oke,fayer}.  This technique has several advantages. It
extends for over five orders of magnitude in time.  The quantity
measured is the derivative of a time correlation function, and
therefore the analog of the distribution of time scales, $\pi(t)$,
discussed above.  And vibrations, which are neglected in our approach,
affect very little the measured decay in the timescales of interest.

In Ref.\ \cite{fayer}, the experimental results were fitted using the
empirical form
\begin{equation}
\pi(t) = [ p t^{-1+C} + d t^{b-1} ] \exp(-t/\tau),
\label{empi}
\end{equation}
with the values $b \approx 0.8 - 0.85$ and $C \approx 0. - 0.2$
consistently found in different liquids.  The $1/t$ behaviour of
$\pi(t)$ obtained at short times when $C=0$ implies a $\log{t}$
behaviour of the time correlator.  This was taken as a challenging
result to mode-coupling theory which does not naturally predict such
patterns, as can be seen when asymptotic analytic results are
considered~\cite{mct}.  These data were however recently virtually
perfectly fitted using a standard schematic version of the
mode-coupling theory, therefore nullifying the criticisms raised in
the experimental work \cite{sperl}.  Note that several fitting
parameters are allowed by the mode-coupling fits, which are performed
using a two-correlator schematic model: coupling constant, distance to
the dynamic singularity and various numerical factors adapting
theoretical time scales to the experimental ones.  By contrast, we
make use of just one free parameter in Fig.~\ref{bpm}.

The `nearly logarithmic decay' of correlations has a simple
explanation in our approach.  At modest temperatures, $T \lesssim
T_o$, there is a coexistence of isolated excitations, responsible for
`slow processes', and rapidly relaxing clusters of excitations; see
Ref.\ \cite{pre} for a detailed discussion of the related temperature
crossovers.  This coexistence manifests as a nearly flat distribution
$\pi(\log t)$ over a wide range of timescales.  This translates into a
$1/t$ behaviour of $\pi(t)$, and logarithmic decay of $P(t)$, as
observed in Fig.~\ref{bpm}.  This explanation, moreover, predicts that
the effective exponent $C$ appearing in Eq.~(\ref{empi}) should
acquire a slight temperature dependence, and change from $C < 0$, when
fast processes dominate closer to $T_o$, to $C>0$ at lower temperature
when slow processes become dominant.  This is precisely what is found
in experiments~\cite{fayer}.  This subtlety is also accounted for by
mode-coupling theory~\cite{sperl}.

The only real discrepancy between fits and data is visible at very
short time, as was anticipated, but the overall quantitative agreement
is very good.  As already suggested by the qualitative analysis of
Ref.~\cite{pre}, our theoretical approach can be used even far above
the experimental glass transition.  Mode-coupling theory is thought to
be applicable in this regime, under the assumption that the dynamics
at modest supercooling is different from that near $T_g$.  As in
Ref.~\cite{pre}, our results here suggest that this may not be the
case.

\begin{figure}
\psfig{file=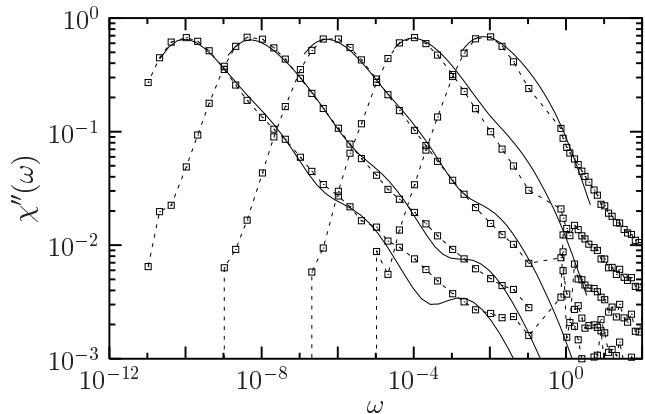,width=8.5cm}
\caption{\label{salol} Dielectric susceptibility data for Salol
obtained from the authors of Refs.~\cite{viot} at temperatures $T=255
K$, $243 K$, $233 K$, $225 K$, and $219 K$ (from right to left). Full
lines are corresponding data for the NEF model at temperatures
$T=0.385$, $0.26$, $0.198$, $0.168$, and $0.15$ (from left to right).
Time is counted in units of 10 ps in the experiments, and in Monte
Carlo units in the simulations.}
\end{figure}

One of the advantages of the present approach on mode-coupling theory
is that it does not produce an unphysical singularity at a temperature
above $T_g$, and it can therefore be used down to very low
temperatures and large relaxation times.  In Fig.~\ref{salol} we use
the NEF model to fit dielectric data taken on Salol \cite{nagel}.  As
before, the fits are done using a single free parameter.
Fig.~\ref{salol} shows that the overall agreement is again very good.
Note in particular that the high-frequency wing is correctly accounted
for by the (discrete) hierarchy of dynamic lengthscales discussed in
the previous sections.

Again, discrepancies due to coarse-graining are evident: absence of
short-time processes, and discreteness of the hierarchy of time
scales.  Discrete scales are particularly evident in the Nagel wing,
where numerical data only produce the `skeleton' of the wing instead
of a smooth curve.  Note that the same experimental data were fitted
in Ref.~\cite{viot} using a frustration limited domain scaling picture
of the glass transition: seven fitting parameters were used there to
obtain a satisfactory fit.  Such data are usually fitted in
experimental papers by empirical forms involving again several free
parameters~\cite{bloch}.

\begin{figure}
\psfig{file=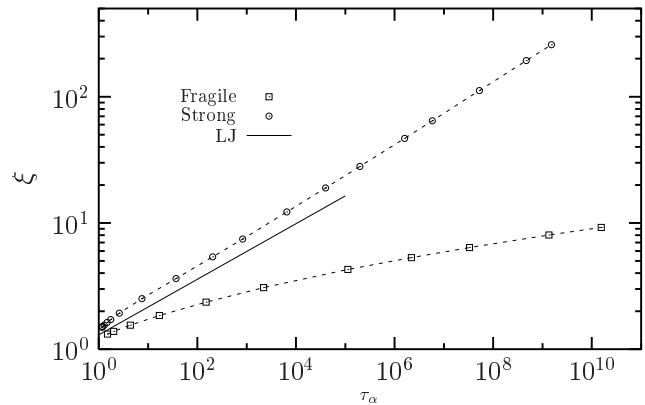,width=8.5cm}
\caption{\label{dyn} Dynamic scaling of timescales versus lengthscales
in the strong three-dimensional FA model~\cite{steve1}, the fragile
three-dimensional NEF model, and three-dimensional binary
Lennard-Jones mixture~\cite{steve1}.}
\end{figure}

Consider, finally, the scaling of lengths and times.  Figure \ref{dyn}
shows dynamic scaling in three different model systems: the
three-dimensional FA model (indicated as `strong'; data from
Refs.~\cite{steve2}), the present NEF model data (indicated as
`fragile'), and the Kob-Andersen Lennard-Jones binary
mixture~\cite{ka} (`LJ' in the figure; data from Ref.~\cite{steve1}).
The figure shows that in the fragile case the growth of dynamic
lengthscales is much less pronounced than in the strong one.  This is
one of the central predictions of the dynamic facilitation approach
\cite{pnas}.  It is a consequence of the temperature dependent dynamic
exponent of East-like models such as the NEF model, in contrast to
strong systems where dynamic lengths go as a fixed power of the
relaxation time.  Note that a figure qualitatively similar to
Fig.~\ref{dyn} would be obtained for the four-point susceptibility as
a function of relaxation time.

This slow growth of lengthscales in the NEF model is also compatible
with the modest size of heterogeneities as suggested by experiments
near $T_g$ \cite{DHreviews2}.  Interestingly, the LJ mixture appears
in this representation as a fairly strong system, despite the common
assumption that it is a good model for fragile liquids.  The
theoretical and numerical findings reported in Fig.~\ref{dyn} offer a
solution to the paradox that experimentally measured length scales are
very small when compared to extrapolations performed from numerical
works~\cite{DHreviews2,DHreviews3}.

\section{Conclusions}
\label{conclusion}

In summary, we have presented an extensive numerical study of the
North-or-East-or-Front model, the three-dimensional generalization of
the East facilitated spin model.  The NEF model, whose dynamical rules
have an externally imposed asymmetry, is in the same universality
class of the more physical arrow model of Ref.\ \cite{pnas} in the
limit of maximal directional persistence, and therefore models the
behaviour of fragile, super-Arrhenius, glass formers.  

We have characterized the equilibrium dynamics of the NEF model
through measurements of the $\alpha$-relaxation timescales,
distributions of relaxation times, and the growth of dynamic
correlation lengths and four-point susceptibilities.  We find that
$\tau_\alpha$ follows a B\"assler law of exponential-inverse-squared
temperature dependence, and that the dynamic correlation length grows
as $\xi \sim c^{-\nu}$ with $\nu \approx 1/3$, where $c$ is the
equilibrium concentration of excitations, while the dynamic
susceptibility grows as $\chi \sim c^{-\gamma}$ with $\gamma \approx
1$.  These results confirm some of the expectations of Ref.\
\cite{pnas} for the arrow model, in particular the quasi
one-dimensional nature of the dynamics in all dimensions.  The decay
of the four-point structure factor, $S(k) \approx k^{-2-1.58}$, is a
direct generalization of the result for the 1D East model
\cite{garrahan-chandler}, and an indication of hierarchical dynamics
in the NEF model.  These two features, persistence of directionality
and hierarchical dynamics, give rise to fragile behaviour \cite{pnas}.

We have shown how the NEF model can be used to rationalize
experimental observations over a wide range of temperatures and
timescales.  We have fitted correlation data in the time domain,
obtained in the mildly supercooled regime by optical Kerr effect
\cite{fayer}, with the NEF model predictions.  This provides an
interpretation of the quasi logarithmic time dependence of
correlations based on the coexistence of isolated excitations and
clusters of excitations, which occurs when the system is crossing over
from a regime of homogeneous dynamics at high $T$ to one of
heterogeneous dynamics at low $T$ \cite{pre}.  We have shown how the
NEF model can also be used to describe dielectric susceptibility data,
in the frequency representation, measured close to the glass
transition by dielectric spectroscopy.  Interestingly, the NEF model
successfully accounts for the excess high-frequency or Nagel wing
\cite{nagel}.

The results presented in this paper add to the list
\cite{pnas,jcp,jung,Berthier-et-al} of phenomenological observations
that can be quantitatively understood within the dynamic facilitation
approach.

\acknowledgments

We are grateful to the authors of Refs.~\cite{bloch,fayer,viot}
for providing us with their published data, and to the authors of
Ref.~\cite{sperl} for sharing their published theoretical results.  We
thank G.~Biroli, J.-P.~Bouchaud, W.~G\"otze, 
G.~Tarjus, and especially D. Chandler
for stimulating 
discussions and comments.  This work was supported by CNRS
(France), EPSRC Grants No.\ GR/R83712/01 and GR/S54074/01, 
and University of Nottingham
Grant No.\ FEF 3024.


\begin{thebibliography}{99}
\bibitem{footnote} Beyond the one-spin facilitated FA model,
Fredrickson and Andersen have also introduced more constrained models,
namely $n$-spin isotropically facilitated models with $n > 1$.  These
models do exhibit fragile behaviour~\cite{fa}, but they are in a sense
not realistic since they are characterized by exponentially diverging
activation energies~\cite{giulio}, a feature not observed in real
liquids.
\end{thebibliography}
\end{document}